\documentclass[runningheads]{llncs}
\usepackage[T1]{fontenc}
\usepackage{graphicx}
\begin{document}
\title{Towards an Inclusive Digital Society: Digital Accessibility Framework for Visually Impaired Citizens in Swiss Public Administration}
\titlerunning{Towards an Inclusive Digital Society}
%
\author{Sabina Werren\inst{1} \and
Hermann Grieder\inst{1}\orcidID{0000-0001-9984-8615} \and
Christopher Scherb\inst{1}\orcidID{0000-0001-6116-5093}}

\authorrunning{S. Werren et al.}
%
\institute{
University of Applied Sciences and Arts, Northwestern Switzerland \\
sabina.werren@students.fhnw.ch, hermann.grieder@fhnw.ch, 
}
\maketitle              
\begin{abstract}
As we progress toward Society 5.0's vision of a human-centered digital society, ensuring digital accessibility becomes increasingly critical, particularly for citizens with visual impairments and other disabilities. This paper examines the implementation challenges of accessible digital public services within Swiss public administration. Through Design Science Research, we investigate the gap between accessibility legislation and practical implementation, analyzing how current standards translate into real-world usability.

Our research reveals significant barriers including resource constraints, fragmented policy enforcement, and limited technical expertise. To address these challenges, we present the Inclusive Public Administration Framework, which integrates Web Content Accessibility Guidelines with the HERMES project management methodology. This framework provides a structured approach to embedding accessibility considerations throughout digital service development.
Our findings contribute to the discourse on digital inclusion in Society 5.0 by providing actionable strategies for implementing accessible public services. As we move towards a more integrated human-machine society, ensuring digital accessibility for visually impaired citizens is crucial for building an equitable and inclusive digital future.
\end{abstract}
\section{Introduction}

The digital age has fundamentally transformed how we interact with government services. As we progress toward Society 5.0, a human-centered society that balances economic advancement with social challenges, ensuring digital accessibility becomes increasingly critical. However, this transformation has created new barriers for individuals with disabilities, particularly those with visual impairments~\cite{kushwaha2024}~\cite{kerdar2024}.

In Switzerland, where digital transformation of public services is rapidly advancing, approximately 61\% of citizens express readiness to engage with digital public services~\cite{deloitte2022}. However, this general acceptance masks significant accessibility challenges. Despite Switzerland's commitment to inclusive accessibility through the Federal Act on the Elimination of Disadvantages for Persons with Disabilities (DDis)~\cite{behig2002}, many digital platforms remain difficult or impossible for visually impaired individuals to navigate independently.

Recent incidents highlight the persistent gap between policy and practice. For instance, in 2023, a mandatory Federal Statistical Office survey was found to be inaccessible to blind individuals~\cite{jaun2023}. Such cases underscore the critical need for better alignment between accessibility legislation and practical implementation. With approximately 1.8 million people in Switzerland reporting disabilities in 2020 (21\% of the population)~\cite{statistik2023}, addressing these accessibility gaps becomes paramount.

This research examines the challenges and opportunities in implementing accessible digital public services within Swiss public administration, focusing on visual impairment. Through Design Science Research, we investigate how current web accessibility standards~\cite{wcag2024} and regulations translate into real-world usability. To address identified barriers, we present the Inclusive Public Administration Framework, which integrates Web Content Accessibility Guidelines (WCAG) with the HERMES project management methodology~\cite{Hermes2005}~\cite{ulrich2024} widely used in Swiss public administration.

The remainder of this paper is organized as follows: Section~\ref{sec:litrev} reviews theoretical foundations, Section~\ref{sec:resmeth} presents our methodology, Section~\ref{sec:dev} describes the framework development, Section~\ref{sec:eval} presents evaluation results, Section~\ref{sec:disc} discusses implications, and Section~\ref{sec:conc} concludes with recommendations for future research and practice.

\section{Literature Review}
\label{sec:litrev}

Digital accessibility has become increasingly critical in the context of Society 5.0, which emphasizes human-centered technological advancement~\cite{kushwaha2024}. The digital divide, initially focused on basic computer access, now encompasses broader issues of digital literacy and participation~\cite{nguyen2022}. According to the World Health Organization, approximately 15\% of the global population lives with disabilities~\cite{who2011}, with Switzerland reporting a higher rate of 21\%~\cite{statistik2023}.

Visual impairment presents unique challenges in digital environments, requiring various accommodations depending on severity~\cite{who2019}. While moderate vision loss may require magnification and contrast adjustments, total blindness necessitates screen readers and specialized assistive technologies~\cite{kerdar2024}. Recent incidents, such as Switzerland's inaccessible Federal Statistical Office survey~\cite{jaun2023}, highlight ongoing gaps between accessibility policies and their practical implementation.

Switzerland's approach to digital accessibility is governed by the Federal Act on the Elimination of Disadvantages for Persons with Disabilities (DDis)~\cite{behig2002}, complemented by international standards like the Web Content Accessibility Guidelines (WCAG)~\cite{wcag2024}. The HERMES project management methodology, widely used in Swiss public administration~\cite{Hermesplatform2005}, has begun incorporating accessibility considerations~\cite{ulrich2024}, though implementation challenges persist.

While digital transformation of public services offers opportunities for increased accessibility, poor implementation can create new barriers. Studies show that 61\% of Swiss citizens are ready for eGovernment services~\cite{deloitte2022}, but this general readiness doesn't reflect varying accessibility needs. Emerging technologies, particularly artificial intelligence, present new possibilities for enhancing accessibility~\cite{hamideh2024}, though their implementation requires careful consideration.

Despite advances in policy and technology, a significant gap remains between theoretical frameworks and practical implementation of digital accessibility in public administration. This gap, shaped by resource constraints, technical limitations, and organizational challenges, forms the focus of our research, which aims to develop a practical framework integrating accessibility considerations into existing project management methodologies.

\section{Research Methodology}
\label{sec:resmeth}

This research employs Design Science Research (DSR) methodology to address the challenges of digital accessibility in Swiss public administration. DSR's focus on creating and evaluating artifacts to solve organizational problems~\cite{hevner2010} aligns with our goal of developing a practical framework for implementing accessibility in public services. The methodology's emphasis on utility and effectiveness~\cite{peffers2007} corresponds with our aim to create actionable solutions, while its iterative approach enables continuous refinement through stakeholder feedback.

In the context of Society 5.0, where human-centered technological solutions are paramount, DSR's focus on creating practical, user-oriented artifacts makes it particularly appropriate. The methodology enables us to systematically address the socio-technical challenges of digital accessibility while ensuring solutions remain grounded in real-world applicability~\cite{vaishnavi2015}.

Our research follows five sequential yet iterative phases:

The Problem Awareness phase began with a systematic literature review of digital accessibility challenges and current standards, complemented by expert interviews with stakeholders to identify practical barriers in implementing accessible digital services. This phase provided a comprehensive foundation for solution development by documenting real-world accessibility challenges within Swiss public services.

During the Solution Suggestion phase, we analyzed interview findings to identify key requirements and reviewed existing frameworks and methodologies. This analysis informed the initial conceptualization of our accessibility framework, ensuring alignment with the HERMES project management methodology commonly used in Swiss public administration.

The Development phase focused on creating the Inclusive Public Administration Framework, carefully integrating WCAG principles with HERMES methodology. The framework underwent continuous refinement based on stakeholder feedback to ensure practical applicability. In the Evaluation phase, we distributed surveys to cantonal project managers to assess the framework's effectiveness, collecting detailed feedback on practicality and implementation feasibility.

Data collection employed multiple methods to ensure comprehensive understanding. We conducted semi-structured interviews with key stakeholders including project managers, heads of eChannels, product owners, and IT providers specializing in cantonal systems. These interviews were analyzed using Mayring's qualitative content analysis approach~\cite{mayring2015}, enabling identification of recurring themes and critical insights. The framework evaluation employed a mixed-methods approach, combining quantitative assessment through structured surveys with qualitative feedback through open-ended questions.

Our literature review encompassed academic publications, policy documents, and technical standards, focusing particularly on accessibility guidelines, legislative frameworks, and developments in assistive technologies. This provided both theoretical foundation and identification of current best practices in digital accessibility.

Several methodological limitations must be acknowledged. The study's focus on cantonal-level administration may limit generalizability to other administrative levels. Additionally, our emphasis on visual impairment means other disabilities received less attention in framework development. Time constraints affected the depth of framework evaluation, potentially limiting observation of long-term implementation effects. These limitations were carefully considered throughout our analysis and interpretation of results, informing our recommendations for future research.

\section{Development of an Inclusive Public Administration Framework}
\label{sec:dev}

The Inclusive Public Administration Framework represents a novel integration of established methodologies with new approaches to accessibility implementation. Our framework builds upon three existing foundational elements: the Web Content Accessibility Guidelines (WCAG)~\cite{wcag2024}, the HERMES project management methodology~\cite{Hermesplatform2005}, and Swiss accessibility legislation~\cite{behig2002}. While these elements provide robust individual foundations, our contribution lies in their innovative integration and extension to address specific challenges in Swiss public administration.

\subsection{Framework Foundations and Novel Integration}
\begin{figure}[ht]
    \centering
    \includegraphics[width=0.7\textwidth]{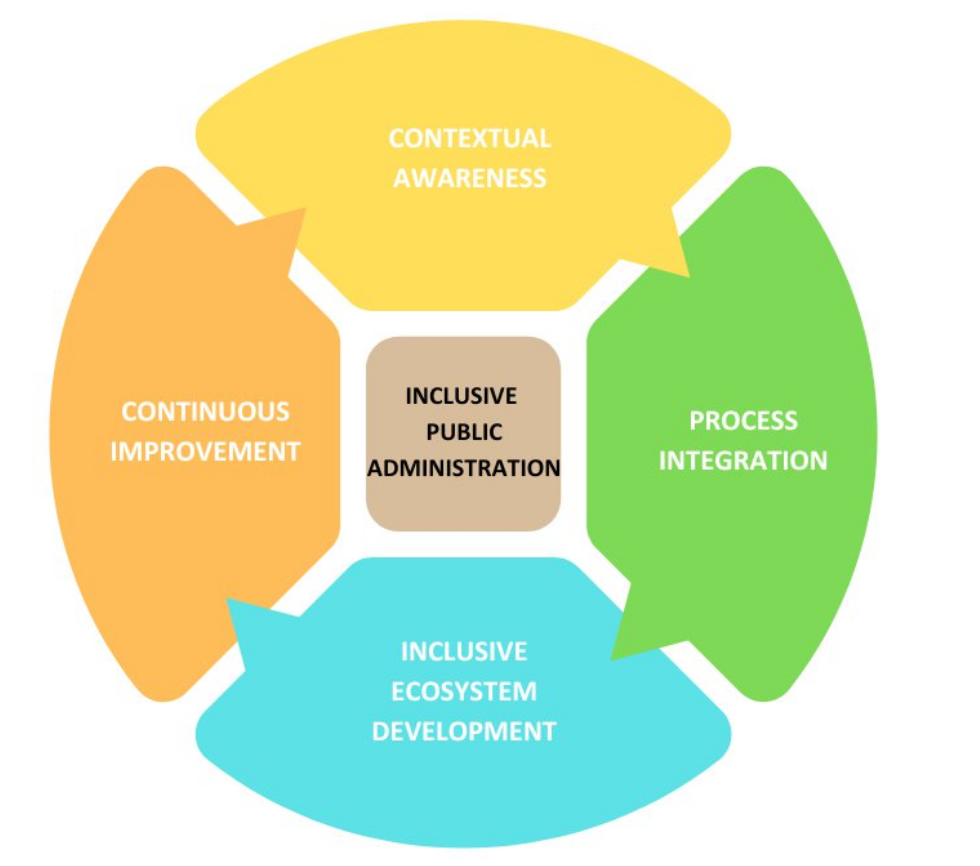}
    \caption{Visualization of the Inclusive Public Administration Framework logic}
    \label{fig:framework-logic}
\end{figure}

The foundational structure of our framework, illustrated in Figure~\ref{fig:framework-logic}, demonstrates how we systematically integrate accessibility considerations throughout the public service development lifecycle. This visualization shows how our approach addresses both website content management and software development projects, ensuring comprehensive coverage of digital public services.

\subsection{Integration with HERMES Methodology}
\begin{figure}[ht]
    \centering
    \includegraphics[width=1\textwidth]{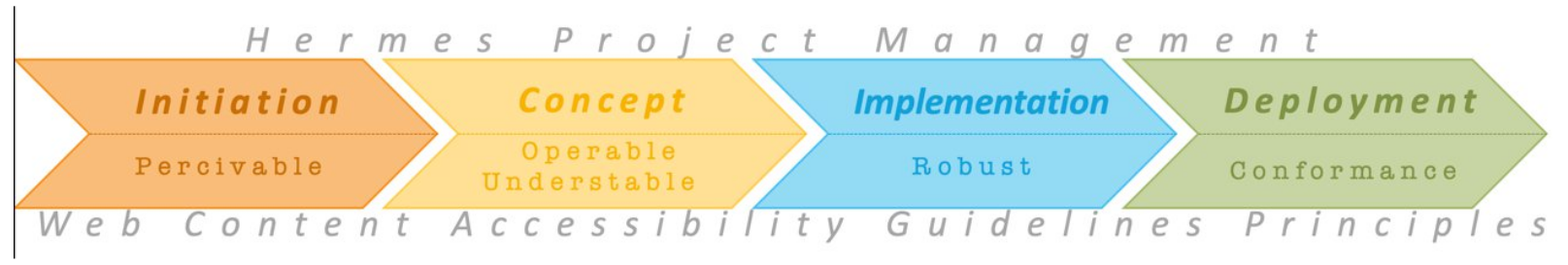}
    \caption{Integration of WCAG principles with HERMES project management methodology}
    \label{fig:framework-hermes}
\end{figure}

As shown in Figure~\ref{fig:framework-hermes}, our framework introduces accessibility-focused elements within each HERMES phase while maintaining the methodology's established structure~\cite{ulrich2024}. This integration creates a seamless blend of project management practices and accessibility requirements, representing a significant contribution to both fields.

\subsubsection{Initiation Phase (Perceivable)}
The initiation phase extends HERMES's traditional project inception activities with novel accessibility considerations. While HERMES typically focuses on general project planning, our framework introduces specific mechanisms for embedding accessibility requirements from the outset. We developed new approaches for integrating WCAG's perceivability principle into initial project documentation and planning processes, creating original templates and guidelines for accessibility-aware project initiation.

\subsubsection{Concept Phase (Operable and Understandable)}
Our innovative approach to the concept phase involves developing new methodologies for translating WCAG's operability and understandability principles into practical project requirements. Our framework systematically addresses key accessibility considerations during this crucial planning stage, ensuring that both technical and user experience aspects are comprehensively covered.

\subsubsection{Implementation Phase (Robust)}
In the implementation phase, we introduce specialized technical implementation guidelines that bridge the gap between WCAG requirements and HERMES development processes. Our newly developed workflows for integrating accessibility testing into existing development practices include specific protocols for assistive technology compatibility testing within the Swiss administrative context.

\subsubsection{Deployment Phase (Conformance)}
The deployment phase structure represents our novel approach to ensuring long-term accessibility maintenance in public services. Our original framework for accessibility auditing and maintenance planning is specifically designed to address the unique challenges of Swiss public administration.

\subsection{Novel Framework Components}
Building upon established accessibility standards, we developed several original components to address specific needs in Swiss public administration:

\begin{enumerate}
    \item \textbf{Customized Accessibility Checklists:} New assessment tools that combine WCAG requirements with HERMES project phases.
    
    \item \textbf{Role-Specific Guidelines:} Original documentation defining accessibility responsibilities within Swiss administrative structures.
    
    \item \textbf{Context-Specific Testing Protocols:} Newly developed testing approaches tailored to cantonal administrative requirements.
    
    \item \textbf{Implementation Templates:} Original templates and tools designed specifically for Swiss public service contexts.
\end{enumerate}

\subsection{Integration with Existing Processes}
While building upon established administrative procedures, our framework introduces new methods for implementing accessibility requirements within existing constraints. We developed original approaches for resource allocation and accessibility implementation that acknowledge and work within the specific limitations of Swiss public administration. This practical integration represents a significant contribution to the field, demonstrating how accessibility requirements can be effectively implemented within existing organizational structures.

These novel contributions collectively create a framework that not only builds upon established methodologies but also introduces new, practical approaches to implementing accessibility in public services. Our work addresses a significant gap in existing practice while providing concrete solutions for accessibility implementation in Swiss public administration.

\section{Evaluation of the Inclusive Public Administration Framework}
\label{sec:eval}

The evaluation of the Inclusive Public Administration Framework was conducted through a comprehensive assessment involving cantonal project managers and stakeholders in Swiss public administration. Our evaluation approach focused on three key dimensions: policy and process alignment, technical usability, and adaptability for future needs. This section presents the evaluation results and their implications for framework implementation.

\subsection{Evaluation Methodology}
The assessment was conducted through a structured survey distributed to project managers involved in digital transformation initiatives at the cantonal level. The survey employed both quantitative measures using a 5-point Likert scale and qualitative feedback through open-ended questions, allowing for a comprehensive understanding of the framework's effectiveness.

\subsection{Policy and Process Alignment}
\begin{figure}[ht]
    \centering
    \includegraphics[width=0.8\linewidth]{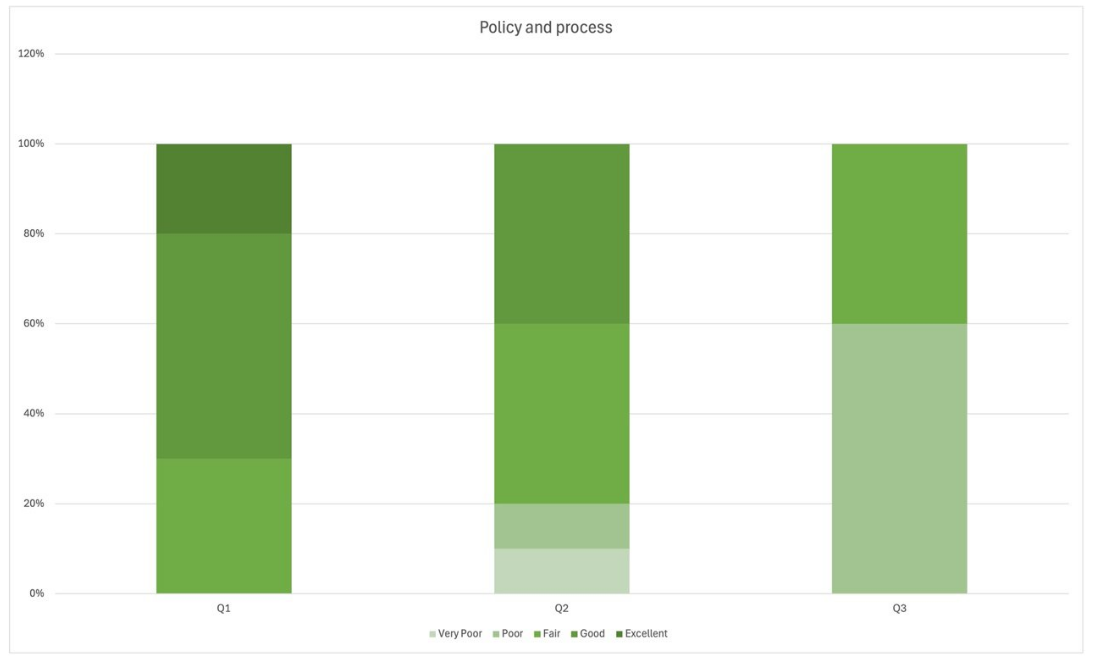}
    \caption{Stakeholder assessment of framework's policy alignment and process integration}
    \label{fig:policy-alignment}
\end{figure}

Our evaluation first examined how effectively the framework aligns with existing internal regulations and policies. Figure~\ref{fig:policy-alignment} presents the stakeholder feedback regarding policy alignment and enforcement gaps.

The results indicate moderate to strong alignment with existing regulations, with 60\% of respondents rating the framework's effectiveness as 'good' or 'excellent.' However, the evaluation also revealed challenges in role definition and responsibility assignment, with 40\% of participants indicating that clearer stakeholder role delineation would be beneficial.

\subsection{Technical Usability Assessment}
\begin{figure}[ht]
    \centering
    \includegraphics[width=0.8\linewidth]{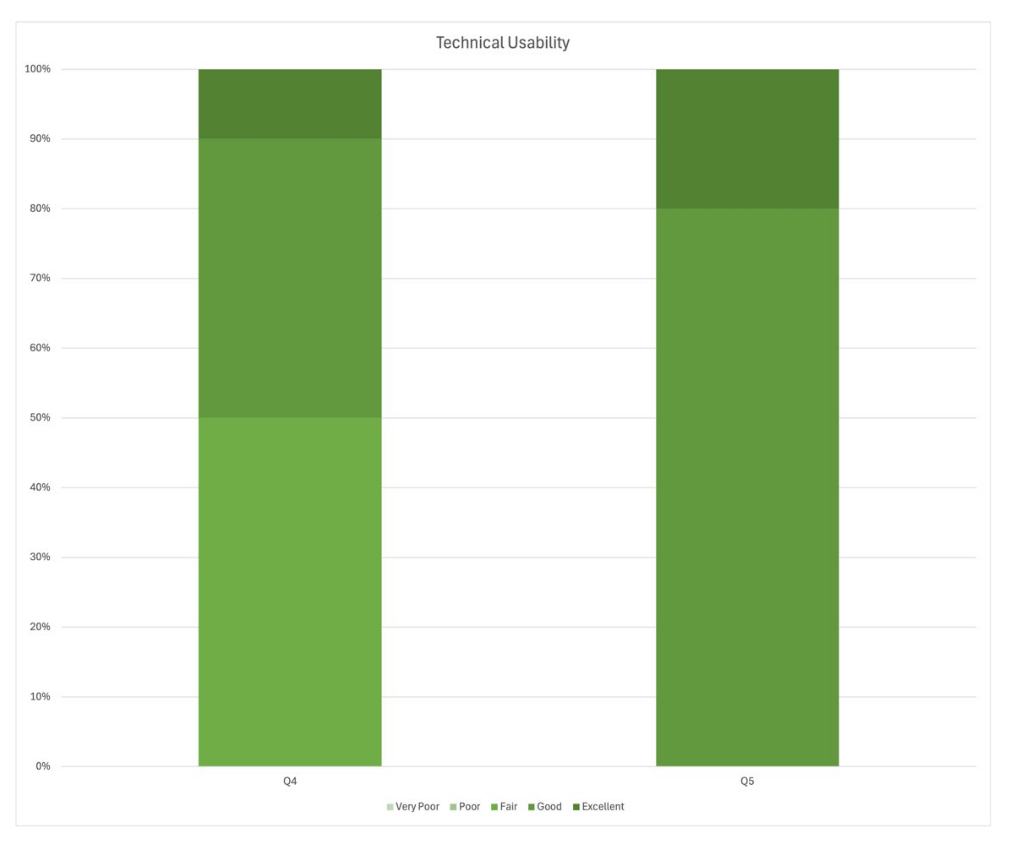}
    \caption{Technical usability assessment results}
    \label{fig:technical-usability}
\end{figure}

The technical assessment focused on the framework's practical implementation aspects and ease of adoption. Figure~\ref{fig:technical-usability} illustrates the stakeholders' evaluation of technical clarity and implementation feasibility.

A significant majority (80\%) of respondents rated the framework's technical guidance as 'good,' indicating strong potential for practical implementation. The evaluation highlighted particularly positive feedback regarding the framework's integration with existing technical processes, with minimal additional training requirements reported.

\subsection{Future Adaptability}
\begin{figure}[ht]
    \centering
    \includegraphics[width=0.8\linewidth]{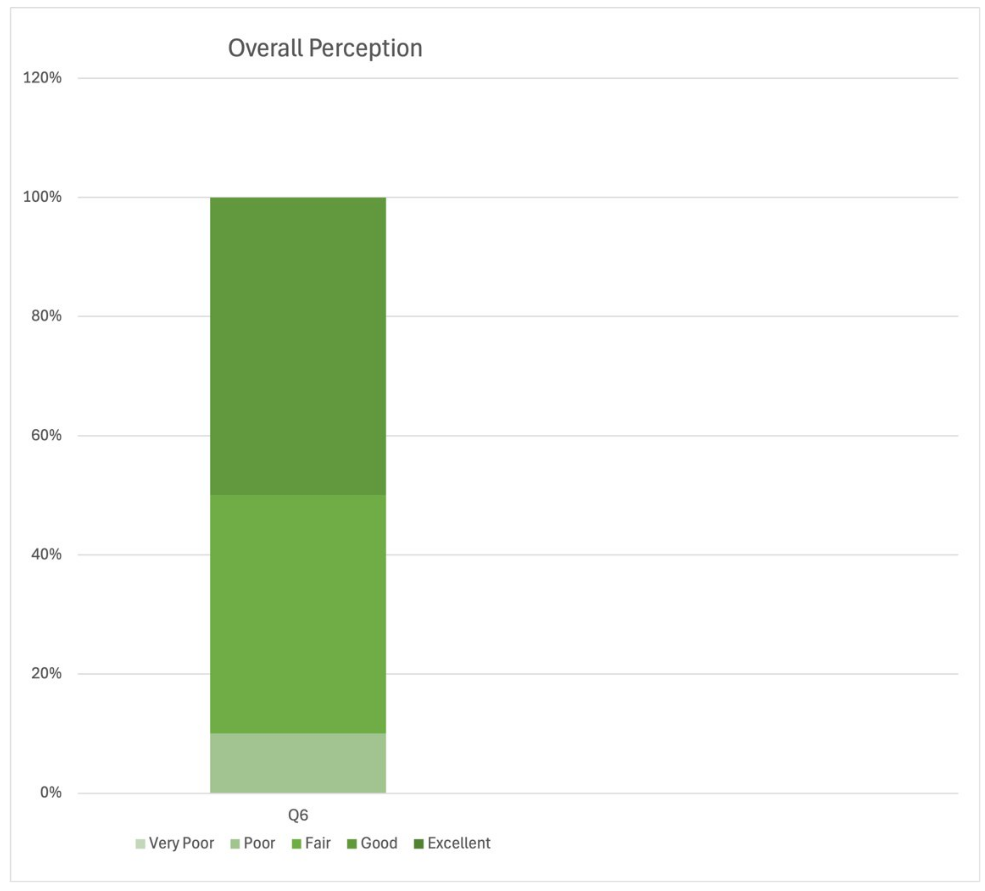}
    \caption{Assessment of framework adaptability for future technologies}
    \label{fig:future-adaptability}
\end{figure}

The framework's capacity to accommodate future technological advancements was assessed through both quantitative ratings and qualitative feedback. Figure~\ref{fig:future-adaptability} presents the stakeholders' perspectives on the framework's adaptability.

The evaluation revealed strong confidence in the framework's adaptability, with 60\% of respondents rating it as 'good' for future technological integration. Stakeholders particularly appreciated the framework's modular structure, which allows for incorporation of emerging accessibility technologies and standards.

\subsection{Implementation Recommendations}
Based on the evaluation results, several key recommendations emerged for successful framework implementation:

\begin{enumerate}
    \item \textbf{Enhanced Role Definition:} Development of more detailed role descriptions and responsibility matrices to address the identified clarity gaps in stakeholder responsibilities.
    
    \item \textbf{Training Support:} Creation of supplementary training materials and guides to support framework adoption, despite the generally positive assessment of implementation ease.
    
    \item \textbf{Customization Guidelines:} Development of specific guidelines for framework adaptation to different organizational contexts within cantonal administration.
    
    \item \textbf{Monitoring Mechanisms:} Establishment of clear monitoring and feedback processes to ensure sustained accessibility compliance.
\end{enumerate}

\subsection{Evaluation Limitations}
Several limitations of our evaluation approach should be acknowledged:

The evaluation primarily focused on cantonal-level administration, potentially limiting insights into framework applicability at other governmental levels. The assessment timeframe provided limited opportunity to observe long-term implementation effects, and the reliance on self-reported data may introduce certain biases in our findings. Additionally, while the response rate was adequate for our analysis, a broader sample size could provide more comprehensive insights into framework effectiveness across different administrative contexts.

\subsection{Future Evaluation Needs}
Our assessment revealed areas requiring further evaluation, particularly regarding long-term implementation effects and adaptation to emerging technologies. Future evaluations should consider:

The framework's effectiveness in different administrative contexts, longitudinal studies of accessibility outcomes, and assessment of the framework's impact on end-user experiences. These additional evaluation aspects would provide valuable insights for ongoing framework refinement and adaptation.

\section{Discussion}
\label{sec:disc}

The development and evaluation of the Inclusive Public Administration Framework reveals significant insights about implementing digital accessibility in public services within the context of Society 5.0. Our research demonstrates both the possibilities and challenges of translating accessibility standards into practical implementation. While WCAG guidelines~\cite{wcag2024} and Swiss legislation~\cite{behig2002} provide robust theoretical frameworks, successful implementation requires careful consideration of organizational contexts and constraints. The framework's integration with HERMES methodology~\cite{Hermesplatform2005} demonstrates how established project management practices can be enhanced to systematically address accessibility needs.

With approximately 21\% of Switzerland's population reporting some form of disability~\cite{statistik2023}, ensuring digital accessibility is not merely a technical requirement but a fundamental aspect of public service provision. The high satisfaction rates with the framework's technical guidance (80\% positive) indicate that public administration organizations can implement comprehensive accessibility measures when provided with appropriate tools and methodologies. However, the identified challenges in role definition suggest that organizational aspects require as much attention as technical implementation.

The framework's emphasis on integrating accessibility throughout the project lifecycle aligns with Society 5.0's vision of human-centered technological advancement~\cite{kushwaha2024}. The strong adaptability ratings (60\% positive) suggest its potential to accommodate future technological developments, particularly as artificial intelligence and other emerging technologies create new opportunities for enhancing accessibility~\cite{hamideh2024}. However, resource allocation remains a critical concern, with organizations struggling to balance accessibility requirements against other project demands.

Looking forward, several areas require attention: longitudinal studies to assess long-term impact, investigation of emerging technologies' integration while maintaining usability, and research into effective methods for knowledge transfer within public administration organizations. The framework's development and evaluation provide insights beyond Swiss public administration, suggesting that successful digital inclusion requires a balanced approach addressing both technical and organizational aspects of accessibility implementation.

These findings collectively indicate that while technical solutions for digital accessibility exist, their successful implementation depends heavily on organizational factors and systematic integration into existing processes. As we progress toward Society 5.0, frameworks that address both technical and organizational aspects of accessibility will become increasingly important for ensuring truly inclusive digital transformation.

\section{Conclusion}
\label{sec:conc}

This research has demonstrated how digital accessibility can be systematically integrated into public administration processes through our Inclusive Public Administration Framework, which bridges theoretical requirements and practical implementation needs. By integrating WCAG principles with the HERMES project management methodology, we have developed a practical approach that has proven both feasible and effective, as evidenced by positive evaluation results in technical implementation (80\%) and future adaptability (60\%). The framework's success suggests that organizations can effectively implement accessibility requirements while maintaining efficient development processes, provided there is sustained commitment to accessibility as a core aspect of digital service delivery. As we progress toward Society 5.0, ensuring digital inclusion becomes increasingly critical, and our research demonstrates that systematic approaches to accessibility implementation can help create more inclusive digital public services. Looking forward, the challenge lies not in technical possibilities but in maintaining organizational commitment to accessibility implementation, making it not just a legal requirement but a fundamental aspect of effective public service delivery in an increasingly digital society.
\newpage

%
%
 \bibliographystyle{splncs04}
 \bibliography{mybib}

\end{document}